\documentstyle{mn} 
\input epsf.sty

\newif\ifAMStwofonts 
 
\ifoldfss 
  \newcommand{\rmn}[1] {{\rm #1}}

  \ifCUPmtlplainloaded \else 
    \NewTextAlphabet{textbfit} {cmbxti10} {} 
    \NewTextAlphabet{textbfss} {cmssbx10} {} 
    \NewMathAlphabet{mathbfit} {cmbxti10} {} 
    \NewMathAlphabet{mathbfss} {cmssbx10} {} 
  \fi 
  \ifAMStwofonts 
    \ifCUPmtlplainloaded \else 
      \NewSymbolFont{upmath} {eurm10} 
      \NewSymbolFont{AMSa} {msam10} 
      \NewMathSymbol{\upi}     {0}{upmath}{19} 
      \NewMathSymbol{\umu}     {0}{upmath}{16} 
      \NewMathSymbol{\upartial}{0}{upmath}{40} 
      \NewMathSymbol{\leqslant}{3}{AMSa}{36} 
      \NewMathSymbol{\geqslant}{3}{AMSa}{3E}

       \let\le=\leqslant 
        
    \fi 
  \fi 
\fi 
 
\ifnfssone 
  \newmathalphabet{\mathit} 
  \addtoversion{normal}{\mathit}{cmr}{m}{it} 
  \addtoversion{bold}{\mathit}{cmr}{bx}{it} 
  \newcommand{\rmn}[1] {\mathrm{#1}}

  \newmathalphabet{\mathbfit} 
  \addtoversion{normal}{\mathbfit}{cmr}{bx}{it} 
  \addtoversion{bold}{\mathbfit}{cmr}{bx}{it} 
  \newmathalphabet{\mathbfss} 
  \addtoversion{normal}{\mathbfss}{cmss}{bx}{n} 
  \addtoversion{bold}{\mathbfss}{cmss}{bx}{n} 
  \ifAMStwofonts 
    \ifCUPmtlplainloaded \else 
      \UseAMStwoboldmath 
      \makeatletter 
      \new@mathgroup\upmath@group 
      \define@mathgroup\mv@normal\upmath@group{eur}{m}{n} 
      \define@mathgroup\mv@bold\upmath@group{eur}{b}{n} 
      \edef\UPM{\hexnumber\upmath@group} 
      \new@mathgroup\amsa@group 
      \define@mathgroup\mv@normal\amsa@group{msa}{m}{n} 
      \define@mathgroup\mv@bold\amsa@group{msa}{m}{n} 
      \edef\AMSa{\hexnumber\amsa@group} 
      \makeatother 
      \mathchardef\upi="0\UPM19 
      \mathchardef\umu="0\UPM16 
      \mathchardef\upartial="0\UPM40 
      \mathchardef\leqslant="3\AMSa36 
      \mathchardef\geqslant="3\AMSa3E 

       \let\le=\leqslant 

    \fi 
  \fi 
\fi 
 
\ifnfsstwo 
  \newcommand{\rmn}[1] {\mathrm{#1}}

  \DeclareMathAlphabet{\mathbfit}{OT1}{cmr}{bx}{it} 
  \SetMathAlphabet\mathbfit{bold}{OT1}{cmr}{bx}{it} 
  \DeclareMathAlphabet{\mathbfss}{OT1}{cmss}{bx}{n} 
  \SetMathAlphabet\mathbfss{bold}{OT1}{cmss}{bx}{n} 
  \ifAMStwofonts 
    \ifCUPmtlplainloaded \else 
      \DeclareSymbolFont{UPM}{U}{eur}{m}{n} 
      \SetSymbolFont{UPM}{bold}{U}{eur}{b}{n} 
      \DeclareSymbolFont{AMSa}{U}{msa}{m}{n} 
      \DeclareMathSymbol{\upi}{0}{UPM}{"19} 
      \DeclareMathSymbol{\umu}{0}{UPM}{"16} 
      \DeclareMathSymbol{\upartial}{0}{UPM}{"40} 
      \DeclareMathSymbol{\leqslant}{3}{AMSa}{"36} 
      \DeclareMathSymbol{\geqslant}{3}{AMSa}{"3E} 

       \let\le=\leqslant 

    \fi 
  \fi 
\fi 
 
\ifCUPmtlplainloaded \else 
  \ifAMStwofonts \else 
    \def\upi{\pi} 
    \def\umu{\mu} 
    \def\upartial{\partial} 
  \fi 
\fi

\title{On possible `cosmic ray cocoons' of relativistic jets } 
 
\author[Micha{\l} Ostrowski] 
       {Micha{\l} Ostrowski \\ 
Obserwatorium Astronomiczne, Uniwersytet Jagiello\'nski, ul. Orla 171, 
30-244 Krak\'ow, Poland (E-mail: mio{\@@}oa.uj.edu.pl)} 
 
\begin{document} 
 
\maketitle 
 
\label{firstpage} 
 
\begin{abstract} 
We consider effects on an (ultra-) relativistic jet and its ambient
medium caused by high energy cosmic rays accelerated at the jet side
boundary. As illustrated by simple models, during the acceleration
process a flat cosmic ray distribution can be created, with gyroradia
for highest particles' energies reaching the scales comparable to the
jet radius or the energy density comparable to the ambient medium
pressure. In the case of efficient radiative losses a high energy bump
in the spectrum can dominate the cosmic ray pressure. In extreme cases
the cosmic rays are able to push the ambient medium off, providing a
`cosmic ray cocoon' separating the jet from the surrounding medium. The
considered cosmic rays provide an additional jet breaking force and lead
to a number of consequences for the jet structure and its radiative
output. In particular the involved dynamic and acceleration time scales
are in the range observed in variable AGNs.
\end{abstract}
 
 
\section{Introduction} 
 
Shock waves are widely considered as sources of cosmic ray particles in
relativistic jets ejected from active galactic nuclei (AGNs). In the
present paper we consider an alternative, till now hardly explored
mechanism involving particle acceleration at the velocity shear layer,
which must be formed at the interface between the jet and the ambient
medium (cf. discussion, in a different context, of the turbulence role
at a tangential discontinuity by Drobysh \& Ostryakov 1998). One should
note that there is a growing evidence of interaction between the jet and
the ambient medium, and formation of boundary layers, both in
observations (e.g. Attridge et al. 1999, Scarpa et al. 1999, Perlman et
al. 1999) and in modelling (Aloy et al. 1999). In the next section we
discuss this acceleration mechanism in some detail. We point out
possible regimes of turbulent second-order Fermi acceleration at low
particle energies, next dominated by the `viscous' acceleration at
larger energies, and by acceleration at the tangential flow
discontinuity at highest energies. In section 3 we shortly consider
highest energies allowed in this model by radiative and/or escape
losses. Then, in section 4, we discuss time dependent spectra of cosmic
rays accelerated at infinite planar flow discontinuity. In the presence
of efficient radiative losses a usually formed flat power-law
distribution is ended with a bump (in some conditions a nearly
mono-energetic spike) followed by a cut-off. Dynamic consequences of
cosmic ray pressure increase at the jet boundary are discussed in
section 5. In particular a cosmic ray cocoon can be formed around the
jet changing its propagation and leading to an intermittent jet
activity. Final remarks are presented in section 6.
 
One should be aware of a partly speculative presentation character of
this paper. Till now, besides casual remarks, the considered complicated
physical phenomenon was hardly discussed in the literature. One can
mention in this respect a discussion of radiation-viscous jet boundary
layers by Arav \& Begelman (1992) and a discussion of possible cosmic
ray acceleration up to ultra-high energies by Ostrowski (1998a). With
the present paper we would like to open the considered physical
mechanism to more detailed modelling and quantitative considerations.

\section{Particle acceleration at the jet boundary}  
 
For particles with sufficiently high energies the transition layer 
between the jet and the ambient medium can be approximated as a surface 
of discontinuous velocity change, a tangential discontinuity (`td'). If 
particles' gyroradia (or mean free paths normal to the jet boundary) are 
comparable to the actual thickness of this shear-layer interface it 
becomes an efficient cosmic ray acceleration site provided the 
considered velocity difference $U$ is relativistic and the sufficient 
amount of turbulence is present in the medium (Ostrowski 1990, 1998a). 
The problem was extensively discussed in early eighties by Berezhko with 
collaborators (see the review by Berezhko 1990) and in the diffusive 
limit by Earl et al.  (1988) and Jokipii et al. (1989). However, till 
now no one considered the situation with highly relativistic flow 
characterized with the Lorentz factor $\Gamma \equiv (1-U^2)^{-1/2} \gg
1$ and, thus, our present qualitative discussion is mostly based on the
results derived for mildly relativistic flows.
 
Any high energy particle crossing the boundary from, say, region I 
(within the jet) to region II (off the jet), changes its energy, $E$, 
according to the respective Lorentz transformation. It can gain or loose 
energy. In the case of uniform magnetic field in region II, the 
successive transformation at the next boundary crossing, II 
$\rightarrow$ I, changes the particle energy back to the original value. 
However, in the presence of perturbations acting at the particle orbit 
between the successive boundary crossings there is a positive mean 
energy change: 
  
$$<\Delta E>~=\, \eta_{\rmn E} \, (\Gamma-1) \, E \qquad . \eqno(2.1)$$
  
\noindent  
The numerical factor $\eta_{\rmn E}$ depends on particle anisotropy at
the discontinuity. It increases with the growing magnetic field
perturbations' amplitude and slowly decreases with the growing flow
velocity. The last factor will be particularly important for large
$\Gamma$ flows. For mildly relativistic flows, in the strong scattering
limit particle simulations give values of $\eta_{\rmn E}$ as substantial
fractions of unity (Ostrowski 1990). For large $\Gamma$ we will assume
the following scaling
 
$$\eta_{\rmn E} = \eta_{\rmn 0} {2 \over \Gamma} \qquad , \eqno(2.2)$$ 
 
\noindent 
where $\eta_{\rmn 0}$ is defined by the magnetic field perturbations'
amplitude at $\Gamma = 2$. In general $\eta_{\rmn 0}$ depends also on
particle energy. During the acceleration process, particle scattering is
accompanied with the jet's momentum transfer into the medium surrounding
it. On average, a single particle with the momentum $p$ transports
across the jet's boundary the following amount of momentum:
  
$$<\Delta p>\, =\, <\Delta p_{\rmn z}>\, =\,  \eta_p \,(\Gamma-1) \, U 
\, p \qquad , \eqno(2.3)$$ 
  
\noindent  
where the $z$-axis of the reference frame is chosen along the flow
velocity and the value of $p$ is given as the one before transmission.
The numerical factor $\eta_p$ depends on scattering conditions near the
discontinuity and in the highly perturbed conditions (in mildly
relativistic shocks) it can reach values being a fraction of unity also.
At large $\Gamma$ we expect $\eta_p \approx \eta_{\rmn E}$.  As a
result, there acts a drag force {\it per unit surface} of the jet
boundary and the opposite force at the medium along the jet, of the
magnitude order of the accelerated particles' energy density.
Independent of the exact value of $\eta_{\rmn E}$, the acceleration
process can proceed very fast due to the fact that average particle is
not able to diffuse -- between the successive energizations -- far from
the accelerating interface. One should remember that in the case of
shear layer or tangential discontinuity acceleration - contrary to the
shock waves - there is no particle advection off the `accelerating
layer'. Of course, particles are carried along the jet with the mean
velocity of order $U/2$ and, for efficient acceleration, the distance
travelled this way must be shorter than the jet breaking length.
 
The simulations (Ostrowski 1990) show that in favourable conditions the 
discussed acceleration process can be very rapid, with the time scale 
given in the observer frame ($\equiv$ the region II rest frame) 
as\footnote{The expression (2.4) and the following discussion is valid 
for an average accelerated particle. A small fraction of external 
particles reflected from the jet can reach a large energy gain, $\Delta 
E / E \sim \Gamma^2$, but these particles do not play a principal role 
in the cosmic ray energy balance.} 
 
$$ \tau_{\rmn td} = \alpha \, {r_{\rmn g} \over c} \qquad, \eqno(2.4) $$ 
 
\noindent 
where $r_{\rmn g}$ is the characteristic value of particle gyroradius in 
the ambient medium. The introduced acceleration time is coupled to
the acceleration length $l_{\rmn td} \sim \alpha r_{\rmn g}$ due to
particle advection along the jet flow. For efficient scattering the
numerical factor $\alpha$ can be as small as $\sim 10$ (Ostrowski 1990).
A warning should be risen in this place. The applied diffusion model
involves particles with infinite diffusive trajectories between the
successive interactions with the discontinuity. Thus reaching stationary
conditions in the acceleration process requires infinite times, leading
to the infinite acceleration time. However, quite flat spectra, nearly
coincident with the stationary spectrum, are generated in short time
scales given by Eq.~2.4 and these distributions are considered in the
present discussion. One may note that in analytic evaluation of
$\tau_{\rmn td}$ for the ultra-relativistic jet, applying Eq-s~(2.1) and
(2.2), large $\Gamma$ factors cancel each other. For the mean magnetic
field $B_{\rmn g}$ given in the Gauss units and the particle energy
$E_{\rmn EeV}$ given in EeV ($1$ EeV $\equiv 10^{18}$ eV)\footnote{Below
we use also another energy units with respective indices $GeV$ and
$TeV$.} the time scale (2.4) reads as
 
$$\tau_{\rmn td} \sim 10^5 \, \alpha \, E_{\rmn EeV} \, B_{\rmn G}^{-1} \quad [s]
\qquad . \eqno(2.5)$$ 
 
Let us remind that in the case of a non-relativistic jet, $U 
\ll c$, the acceleration process is of the second-order in $U/c$ and a 
rather slow one. 

For low energy cosmic ray particles the velocity transition zone at the
boundary is expected to appear as a finite-width turbulent shear layer.
We do not know of any attempt in the literature to describe the internal
structure of such layer on the microscopic level. Therefore, we limit
the discussion of the acceleration process within such a layer to
quantitative considerations only. From rather weak radiation and the
observed effective collimation of jets in the powerful FR~II radio
sources one can conclude, that the interaction of presumably
relativistic jet with the ambient medium must be relatively weak. Thus
the turbulent transition zone at the jet boundary must be limited to a
relatively thin layer. Within such a layer two acceleration processes
take place for low energy particles (by `low energy particles' we mean
the ones with the mean radial free path $\lambda$ much smaller than the
transition layer thickness, $D$). The first one is connected with the
velocity shear and is called `cosmic ray viscosity' (Earl et al. 1988).
The second is the ordinary Fermi process in the turbulent medium. The
acceleration time scales can not be evaluated with accuracy for these
processes, but -- for particles residing within the considered layer --
we can give an acceleration time scale estimate
 
$$\tau_{\rmn II} = {r_{\rmn g} \over c} {c^2 \over V^2 + \left( U 
{\lambda \over D} \right)^2 } \qquad , \eqno(2.6) $$ 
 
\noindent 
where $V$ is the turbulence velocity ($\sim$ the Alfv\'en velocity for 
subsonic turbulence) and $D$ is the shear layer thickness. The first 
term in the denominator represents the second-order Fermi process, while 
the second term is for the viscous acceleration. One expects that the 
first term can dominate at low particle energies, while the second for 
larger energies, with $\tau_{\rmn II}$ approaching the value given in 
Eq.~(2.4) for $\lambda \sim D$. If the second-order Fermi acceleration 
dominates, $\lambda < D (V/U)$, the time scale (2.6) reads as 
 
$$\tau_{\rmn II} \sim 10^7 \, E_{\rmn TeV} \, B_{\rmn G}^{-1} V_3^{-2}
\quad [s] \qquad , \eqno(2.7)$$ 
 
\noindent 
where $V_3$ is the turbulence velocity in units of $3000$ km/s. 
Depending on the choice of parameters this scale can be comparable or 
longer than the expansion and internal evolution scales for relativistic 
jets. In order to efficiently create high energy particles for the 
further acceleration by the viscous process and the tangential 
discontinuity acceleration one have to assume that the turbulent layer 
includes high velocity turbulence, with $V_3$ reaching values 
substantially larger than $1$. Then the scale (2.7) may be much reduced, 
also because of oblique shocks formed in the turbulent layer and the 
accompanying first order Fermi acceleration processes. For the following 
discussion we will assume that such effective pre-acceleration takes 
place, but the validity of this assumption can be estimated only {\it a 
posteriori} from comparison of our conclusions with the observational 
data. Another possibility is that a population of high energy particles 
exist in the medium surrounding the jet due to some other unspecified 
acceleration processes in the central object vicinity. 
 
The cosmic ray energy spectra generated  with the above mechanisms at
work are expected to be very flat (Section 4; see also Ostrowski 1998a).
With such particle distribution the dynamic influence at the jet and
its' ambient medium can be due to effects of the highest energy cosmic
rays, immediately preceding the spectrum cut-off. Because of the short
acceleration time scale (2.4) expected for such particles the
acceleration process can provide particles (protons) with energies
reaching ultra high energies. Without radiative losses one can obtain
particles with $r_{\rmn g} \sim $ the jet radius, $R_j$~, near the
cut-off in the spectrum (Ostrowski 1998a). For standard jet parameters
the considered particle energies may reach $\sim 10^{19}$ eV. Let us
note that the existence of such cosmic rays was suggested by Mannheim
(1993) to explain $\gamma$-ray fluxes from blazars.
 
\section{Energy losses} 
 
To estimate the upper energy limit for accelerated particles, at first 
one should compare the time scale for energy losses due to radiation and 
inelastic collisions to the acceleration time scale. The discussion of 
possible loss processes is presented by Rachen \& Biermann (1993). The 
derived loss time scale for protons can be written in the form 
    
$$T_{\rmn loss} \simeq 5\cdot 10^9~B_{\rmn G}^{-2} \, (1+Xa)^{-1} \,
E_{\rmn EeV}^{-1} \quad [s] \qquad , \eqno(3.1)$$ 
    
\noindent    
where $B_{\rmn G}$ is the magnetic field in Gauss units, $a$ is the
ratio of the energy density of the ambient photon field relative to that 
of the magnetic field and $X$ is a quantity for the relative strength of 
{\it p}$\, \gamma$ interactions compared to synchrotron radiation. For 
cosmic ray protons the acceleration dominates over the losses (Eq-s~2.5, 
3.1) up to the maximum energy 
    
$$E_{\rmn EeV} \approx 2 \cdot 10^2 \alpha^{-1} \left[ B_{\rmn G} (1+Xa)
\right]^{-1/2} \qquad . \eqno(3.2)$$ 
    
\noindent  
This equation can easily yield a large limiting $E_{\rmn EeV} \sim 1$ 
with moderate jet parameters (e.g. $B_{\rmn G} \approx 1$, $Xa \sim
10^2$, and $\alpha = 10$). However, one should note that the particle gyroradius provides 
the minimum scale for the acceleration region's spatial extent (Ostrowski 1998a). 
Thus, for the actual particle maximum energy $E_{\rmn max}$ the jet 
radius should be larger than the respective particle gyroradius $r_{\rmn 
g}(E_{\rmn max})$. E.g., for $R_j = 10^{16}$ cm and $B_{\rmn G} = 1$ the
particle energy satisfying the condition $R_j = r_{\rmn g}$ equals
$E_{\rmn max}\sim 10$ EeV, what is consistent with the above estimate
based on Eq.~3.2~.

\section{Energy spectra of accelerated particles}
 
The acceleration process acting at the tangential discontinuity of the
velocity field leads to the flat energy spectrum and the spatial
distribution expected to increase their extension with particle energy.
Below, for illustration, we propose two simple acceleration and
diffusion models describing these features. For low and high energy
particles we consider the time dependent acceleration process at,
respectively, the {\em plane} shear layer or tangential discontinuity,
surrounded with infinite regions for particle diffusion. In the
discussion below all particles are ultra-relativistic with $E = p$.

\subsection{A turbulent shear layer}

At first we consider `low energy' particles wandering in an extended
turbulent shear layer, with the particle mean free path $\lambda \propto
p$. With the assumed conditions the mean time required for increasing
particle energy on a small constant fraction is proportional to the
energy itself, and the mean rate of particle energy gain is constant,
$<\dot{p}>_{\rmn gain}$ = const. Let us take a simple expression for the
synchrotron energy loss, $<\dot{p}>_{\rmn loss} \, \propto p^2$, to
represent any real process acting near the discontinuity. With
$<\dot{p}> \, \equiv \, <\dot{p}>_{\rmn gain} - <\dot{p}>_{\rmn loss}$
the transport equation for the particle momentum distribution function
$n \equiv n(t, p, x)$ has the following form
   
$${\partial n \over \partial t} + {\partial \over \partial
p} \left[ <\dot{p}> n \right] + {\partial \over \partial x} \left[
\kappa_\perp {\partial n \over \partial x} \right] + \left( {\partial n
\over \partial t} \right)_{\rmn esc} = Q \quad . \eqno(4.1)$$
   
\noindent   
where $x$ measures the distance perpendicular to the shear layer and the
escape term at highest energies is represented by $\left(
{\partial n \over \partial t} \right)_{\rmn esc}$. 

For the jet boundary acceleration, the jet radius and the escape
boundary distance provide energy scales to the process. Another scale
for particle momentum, $p_{\rmn c}$, is provided as the one for equal
losses and gains, $<\dot{p}>_{\rmn gain} = <\dot{p}>_{\rmn loss}$. As a
result, a divergence from the power-law and a cut-off have to occur at
high energies in the spectrum. At small energies, where the diffusive
regions are extended and losses non-significant, the considered solution
should be close to the power-law.
 
We used a simple Monte Carlo simulations of the acceleration process to
solve Eq.~(4.1). In the equation we assume a continuous particle
injection, uniform within the considered layer, $Q =$ const. The
diffusion coefficient $\kappa_\perp$ is taken to be proportional to
particle momentum, but independent of the spatial position $x$. With
neglected particle escape through the shear layer side boundaries and
the considered uniform conditions ${\partial n \over \partial x} = 0$ and
the spatial diffusion term in Eq.~4.1 vanishes. For the escape term
$\left( {\partial f \over \partial t} \right)_{\rmn esc}$ we simply
assume a characteristic escape momentum $p_{\rmn max}$. At figures 1 and
2 we use $p_{\rmn c}$ as a unit for particle momentum, so it defines
also a cut-off for $p_{\rmn c} < p_{\rmn max}$~. At Fig.~1, at small
momenta the spectrum has a power-law form -- in our model $n(t,p)
\propto p^{-2}$ -- with a cut-off momentum growing with time. However,
at long time scales, when particles reach momenta close to $p_{\rmn c}$,
losses lower the value of $<\dot{p}>$ leading to spectrum flattening and
pilling up particles at $p$ close to $p_{\rmn c}$. Then, a low energy
part of the spectrum does not change any more and only a narrow spike at
$p \approx p_{\rmn c}$ grows with time. Let us also note that in the
case of efficient particle escape, i.e. when $p_{\rmn max} < p_{\rmn
c}$, the resulting spectrum would be similar to one of the short time
spectra in Fig.~1, with a cut-off at $\approx p_{\rmn max}$ (cf. Ostrowski
1998a).

\begin{figure} 
 \vspace{6cm} 
 \includegraphics{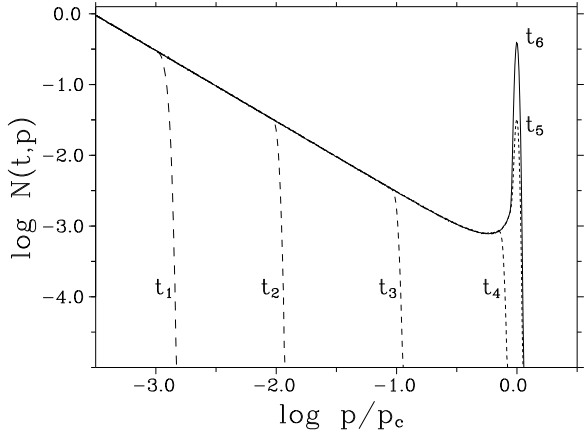} 
 \vspace{0mm} 
 \caption{Spectra $N(t, p) \equiv d\, n / d\, \log\, p$ of accelerated
particles within the boundary shear layer. The results are given in a
sequence of times $t_1 < t_2 < ... < t_6$ ($t_{\rmn i+1} = 10 t_{\rmn
i}$).}
\end{figure}

\subsection{Tangential discontinuity acceleration}

An illustration of the acceleration process at the tangential
discontinuity have to take into account a spatially discrete nature of
the acceleration process. Here, particles are assumed to wander subject
to radiative losses outside the discontinuity, with the mean free path
proportional to particle momentum $p$ and the loss rate proportional to
$p^2$. At each crossing the discontinuity a particle is assumed to gain
a constant fraction $\Delta$ of momentum (cf. Eq-s~2.1, 2.2):

$$p^\prime = (1+\Delta) p \qquad , \eqno(4.2)$$

\noindent
and, due to losses, during each free time $\Delta t$ its momentum
decreases from $p_{\rmn in}$ to $p$ according to the relation

$${1 \over p} - {1 \over p_{\rmn in}} = {\rm const} \cdot \Delta t
\qquad . \eqno(4.3)$$

\noindent
The time dependent energy spectra obtained within this model are
presented in Fig.~2, where we choose units in a way to put constant in
Eq.~(4.3) equal to one and the particle mean free path equals at two
considered models at $p = p_{\rmn c}$. Comparison of the results in two
models allows to evaluate the modification of the acceleration process
by changing the momentum dependence of the particle diffusion
coefficient. For slowly varying diffusion coefficient (represented here
with a `$\lambda = const$' model) high energy particles which diffuse far
away off the discontinuity and loose there much of their energy still
have a chance to diffuse back to be accelerated at the discontinuity. In
the model with $\kappa$ quickly growing with particle energy (here the
`$\lambda = C \cdot p$' model) such distant particles will decrease
their mobility in a degree sufficient to break, or at least to limit
their further acceleration. One should note that in both models the
spectrum inclination at low energies is the same (here $n(p) \propto
p^{-2}$).

\begin{figure} 
 \vspace{11cm}
 \includegraphics{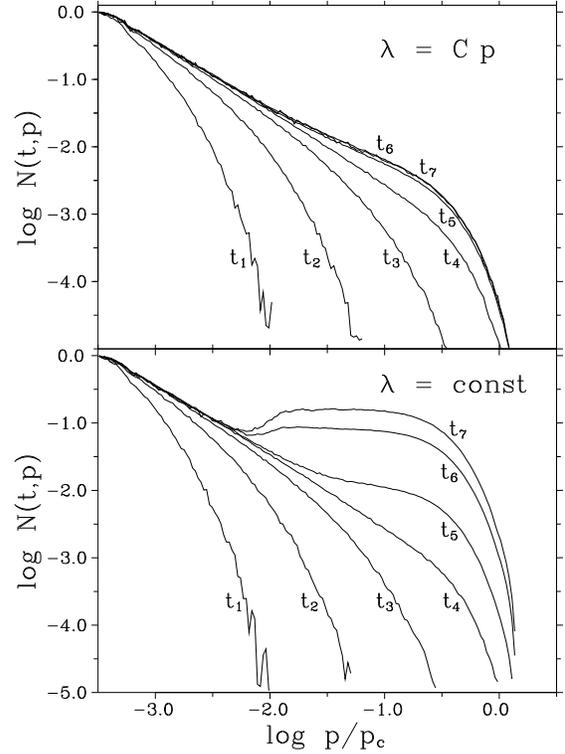}
 \vspace{0mm} 
 \caption{Spectra $N(t, p, x = 0 ) \equiv d\, n / d\, \log\, p$  of
accelerated particles at the jet boundary in a sequence of times $t_1 <
t_2 < ... < t_7$ (each $t_i = 10 t_{i-1}$), for continues particle
injection at small momentum $p_{\rmn 0} << p_{\rmn c}$. The results for
the particle mean free path $\lambda \propto p$ are given in the upper
panel, and for $\lambda = {\rm const}$ in the lower one. A wavy
small momentum feature seen in the spectra is due to injection.}
\end{figure}

\section{Consequences of the jet's `cosmic ray cocoon'} 
 
Cosmic ray distributions in Fig-s~1 and 2 reveal a few spectral
components: a flat power-law section at small energies, followed either
with a smooth transition to the cut-off, or at first a hard component
(`bump' or `spike') proceeding the final cut-off. The later case occurs
when the radiative losses are enough efficient to pile up particles
before the loss dominated high energy range. Spectra without such hard
component will appear in cases when particles escape from the
acceleration region at low energies, or when the acceleration time scale
is longer than the involved dynamical scales (for the jet expansion or
slowing down, e.g.). One may note that in the models discussed by us the
power-law section of the spectrum has the form $n(p) \propto p^{-2}$.

Normalization of the spectrum at low momenta essential for dynamical
considerations is defined by the injection efficiency. This parameter
can not be derived from available models or observations and it is
treated as a free parameter in the present considerations. The cosmic
ray pressure at the jet boundary, $P_{cr} = \int p \, n(p) \, dp$, grows
with growing injection efficiency and extension of the spectrum in
energy. Additionally, the high energy bump can substantially contribute
to $P_{cr}$. Let us review a few possibilities arising due to cosmic ray
population at the jet boundary, as illustrated at Fig.~3~.

Dynamical effects caused by cosmic rays depend on the ratio of
$P_{cr}$ to the ambient medium pressure, $P_{ext}$, at the boundary. If,
at small particle energies, the acceleration time scale is longer than
the jet expansion time or particle escape is efficient at small
energies, the formed energetic particle population cannot reach
sufficiently high energy density to allow for dynamic effects in the
medium near the interface. Then, it acts only as a small viscous agent
near the boundary, decreasing slightly gas and magnetic field
concentration (cf. Arav \& Begelman 1992). In such cases we call the
occurring cylindrically distributed  cosmic ray population a `{\it weak}
cosmic ray cocoon'. Then, if accelerated particles are electrons or can
transfer energy to electrons, a uniform cosmic ray electron population
may be formed along the jet leading to the observed synchrotron
component with slowly varying spectral index and break frequency.
Density of such radiating electrons is expected to have maximum in a
cylindrical layer at the jet boundary.  

If acceleration dominates losses at small (injection) energies, then the
time-dependent high energy part of the spectrum can bear a power-law
form with a growing cut-off energy, like the short time distributions at
Fig-s 1 and 2. After losses become significant a few possibilities
appear. If it happens at low energies, when the acceleration process is
limited to the turbulent shear layer, a power-law with a growing sharp
spike preceding a cut-off energy will appear. If in such case the
increasing cosmic ray pressure in a cocoon could reach values comparable
to the medium pressure, a substantial modification of the jet boundary
layer is expected. Below we will discuss various possibilities arising
in such cases of the `{\it dynamic} cosmic ray cocoons'. If the
acceleration at the tangential discontinuity resembles our models in
section 4.2, then, depending on the conditions near the jet, the cosmic
ray pressure may stabilise at an intermediate $P_{\rmn cr} < P_{\rmn
ext}$, or grow to form the dynamic cosmic ray cocoon.
 
\begin{figure} 
 \vspace{8cm} 
 \includegraphics{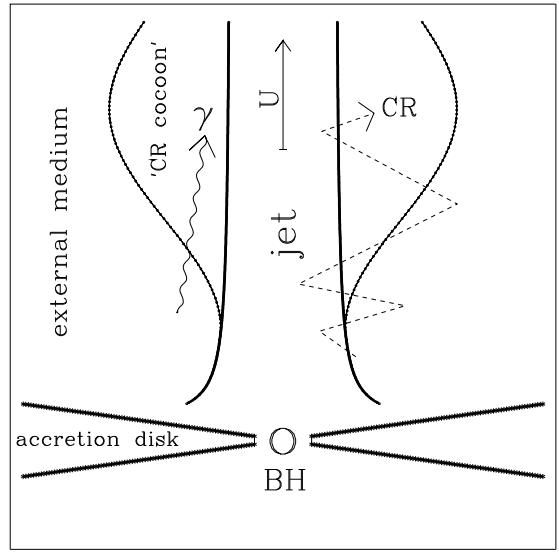} 
 \vspace{0mm} 
 \caption{A schematic view of the situation considered near the central 
source: `BH' is a central black hole, `CR' denotes a cosmic ray particle 
accelerated at the jet boundary, $\gamma$ is a photon radiated from the 
central source vicinity.} \end{figure} 
 
Let us consider a possible scenario of dynamic interaction of high
energy cosmic rays with the jet and the ambient medium. The particles
are `injected' and further accelerated at the jet boundary. Growing
number of such particles results in forming the cosmic ray pressure
gradient outside the jet pushing the ambient medium apart. Additionally,
an analogous gradient may be formed directed into the jet, helping to
keep it collimated. The resulting rarefied medium or partly emptied of
the magnetized plasma space near the jet boundary will decrease the
acceleration efficiency. Thus the cosmic ray energy density may build up
only to the value comparable to the ambient medium pressure, when it is
able to push the magnetized plasma away. Because the diffusive escape of charged particles from the
cosmic ray cocoon is not expected to be efficient (contrary to photons
considered by Arav \& Begelman 1992), in some cases the blown out
volumes could be quite large, reaching the values comparable to $R_j$ or
even to the local vertical scale of gas. The accumulated cosmic rays can
be removed by advection -- in the form of cosmic rays' filled bubbles or
cosmic ray dominated winds -- outside the active nucleus into regions of
more tenuous plasma, or simply outside the jet at larger distances from
the central source.
 
The jet moving in a space filled with photons and high energy cosmic
rays (cf. Fig.~3) is subject to the braking force due to scattering this
species (e.g. Sikora et al. 1998; for the photon breaking). For cosmic
rays with $\lambda \sim R_j$ both types of particles penetrate
relatively freely inside the jet and the breaking force is exerted more
or less uniformly within its volume, in rough proportion to the electron
(or pairs') density for the photon breaking and to the turbulent
magnetic field energy density for the cosmic ray breaking. If cosmic ray
cut-off energy is lower, with the equivalent $\lambda < R_j$, the cosmic
ray breaking force acts within the jet boundary layer of width
$\lambda$. From Eq.~(2.3) we estimate the cosmic ray breaking force {\it
per unit jet length} to be
 
$$ f_{\rmn b,cr} = 2\pi \, \eta_{\rmn  p} \, (\Gamma-1) \, P_{\rmn cr} 
\, R_j \quad , \eqno(5.1)$$ 
 
\noindent 
where we consider  $\lambda \le R_j$ and we put $U=1$. From the above 
discussion, in the stationary conditions one can put $P_{\rmn cr} 
\approx P_{\rmn ext}$, where $P_{\rmn ext}$ is the external medium 
pressure. For a jet with a (relativistic) mass density $\rho_j$, with 
Eq-s~(2.2,3) and $\eta_{\rmn  p} = \eta_{\rmn  E}$, the jet breaking 
length due to cosmic rays is 
 
$$L_{\rmn b,cr} = R_j \, {\Gamma \over 4 \eta_{\rmn 0}} \, {\rho_j c^2 
\over P_{\rmn ext} } \quad . \eqno(5.2) $$ 
 
\noindent 
For example, assuming $P_{\rmn ext} = \rho_j c^2$ and $\eta_{\rmn 0} = 
0.25$, we obtain $L_{\rmn b,cr} = \Gamma R_j$. 
 
Because of dynamic (`d') form of pushing out the ambient medium and 
following it cosmic rays' escape, the back-reaction of this process at 
particle acceleration is expected to make the full process unstable, 
with an intermittent behaviour seen in longer time scales. The full 
configuration with the `heavy' ambient gas supported with the `light' 
gas of ultra-relativistic particles in the cosmic ray cocoon is expected 
to be subject to the Raileigh-Taylor (`RT') instability. A related 
characteristic time scale can be roughly estimated as

$$t_{\rmn RT} \sim \left( {L \over 2 \pi g} \right)^{1/2} \quad ,
\eqno(5.3)$$
 
\noindent 
where $g$ is the gravitational acceleration and $L$ the scale of 
instability (e.g. for $g = 10^2$ cm/s$^2$, $L = 10^{17}$ cm and 
additional requirement of the sound velocity comparable to $c$ the time 
scale $t_{\rmn RT}$ is estimated to be below $1$ yr). 
 
Another type of instability can be generated by the time dependence of 
the non-linear acceleration process. Continues injection of (low energy) 
seed particles to the acceleration process can be continued till the 
cosmic ray pressure becomes equal to $P_{\rmn ext}$. Then, the ambient 
plasma is pushed away from the jet and its interaction with the jet 
boundary surface diminishes. It must lead to decrease of the injection 
efficiency, if acceleration at the turbulent surface layer is 
responsible for the process. Then, cosmic ray energy density contained 
in still accelerated highest energy particles increases until they 
manage to escape from the jet vicinity, allowing for re-establishment of
original conditions. It allows the ambient medium to `fall down' at the
jet to start a new phase of intensive interaction between the jet and
the ambient magnetized plasma, initiating efficient injection of low
energy seed particles. The process should be accompanied with intense
kinetic energy dissipation processes and a radiation flare at all
frequencies. In the flare phase one can expect substantial weakening of
the cosmic ray jet breaking mechanism allowing for larger jet velocity
and forming internal shocks. Next, the full process could repeat with a
time scale comparable to the time required for removing highest energy
particles from the system. For the inequality $t_{\rmn RT} > \tau_{\rmn
td} (E_{\rmn max})$ a continues (diffusive, or as a wind) particle
escape will govern the process. Then one may expect smaller variations
of the output radiation flux.
 
The presented discussion assumed the cylindrical symmetry of the 
unstable flow, which may be not true. However, any {\it large} amplitude 
perturbation of the conditions near/in the jet can not be much smaller 
than the spatial scale $R_j$ and the respective observer's time scale 
shorter than $R_j/c$. 
 
\section{Conclusions and further speculations} 
 
Limited to the hydrodynamic approach the present discussion is intended
to provide an alternative view of the AGN central activity related to
the jet outflows. The acceleration of cosmic rays up to extremely high
energies occurs in a natural way at the relativistic jet boundary if
there are effective preliminary acceleration mechanisms providing seed
particles with mean free paths comparable to the width of a boundary
layer. With the assumption that such processes work efficiently we
discussed several possible consequences for the conditions in the
spatial volume containing the jet. Here, the main factor playing the
role is a flat spectrum cosmic ray population carrying substantial
energy density in highest energy particles. These particles may
dynamically influence the jet flow and conditions in the surrounding
medium, without {\it direct} radiative effects. A possibility is
considered of the jet timely separated from the ambient gas by the layer
filled with cosmic rays and ambient photons. During such a phase the
electromagnetic radiation produced in the jet can more easily escape
from the AGN centre, to reach observer situated close to the jet axis
direction. Also, the plasma self-absorption frequency can be decreased
for such observer if the optical depth of the upper plasma layers does
not dominate the output. As we expect larger intensity of generated
low-frequency radiation when the ambient medium directly pushes on the
jet, the discussed picture should be characterized with a positive
correlation of the radiation intensity with the self-absorption
frequency shift to larger values (see B\"ottcher 1999 for a recent
discussion of such shifts within the standard jet picture). Also, for
flares in some BL Lac objects with a week-month time scale, the
beginning of the flare should be seen approximately at the same time at
all non-absorbed frequencies. Only later evolution of the introduced
disturbance will lead to shifts of flare maxima at different
frequencies. One should remember that we do not include into this
discussion other radiation sources (accretion disc, corona) in the
active nucleus vicinity.
 
The instabilities related to the discussed process may lead to temporary
variations of jet flow velocity and the degree of the jet surface
perturbation. Perturbations in jet flow introduced by these
instabilities may also lead to shock wave formation with its
observational consequences. As a result the processes accelerating lower
energy cosmic rays and cosmic ray electrons are expected to have
fluctuating nature with time scales  estimated in Eq.~(5.3) for large
changes. The processes occurring inside the jet can be characterized
with the observer's time scale a factor of $\Gamma$ shorter. However, in
the situation with the jet perturbation introduced by the external
process, the actual time scale will be intermediate between the
internal, Lorentz contracted one and the external perturbation scale.
 
In the above discussion we avoided considering the acceleration of
electrons (or pairs). In the mentioned model of Mannheim (1993)
energetic electrons arise as a result of cascading of pairs resulting
from the energetic proton interactions inside the jet. One can consider
also different scenarios providing the cosmic ray electrons. E.g.,
in the space close to the jet boundary a large
power can be stored in the highly anisotropic population of cosmic ray
protons. Such distribution is known to be unstable and it leads to
creation of long electromagnetic plasma waves. Damping of such waves by
pairs may be very efficient acceleration process providing cosmic ray
electrons (cf. Hoshino et al. 1992, in a different context). Of course,
the short plasma waves at the jet boundary can be also generated by
velocity shear.
 
In the presented evaluations we often consider the situation with
particles starting to play a dynamic role in the system when their
energies reach scales yielding gyroradia $r_{\rmn g} \sim R_j$. If
particles become dynamically important at lower energies, with $r_{\rmn
g} << R_j$, all considered time scales should be respectively scaled
down. Then, the jet breaking force due to cosmic rays is acting only at
the external layers of the jet, generating magnetic stresses along it.

\section*{Acknowledgements} 
 
Discussions with Mitch Begelman and Marek Sikora were particularly
useful during preparation of this paper. Critical remarks of Luke Drury
helped to improve the final version of the paper. I also gratefully
acknowledge support from the {\it Komitet Bada\'n Naukowych\/} through
the grant PB 179/P03/96/11~, and, partly, within the project
2~P03D~002~17.

 
\label{lastpage} 
 
\end{document}